# Linear Arrays of Metal-Coated Microspheres: a Polarization –Sensitive Hybrid Colloidal Plasmonic-Photonic Crystal

C. Farcau[1,2]

[1]National Institute for Research and Development of Isotopic and Molecular Technologies INCDTIM, Cluj-Napoca, Romania
[2]Institute for Interdisciplinary Research in Bio-Nano-Sciences, Babes-Bolyai University, Cluj-Napoca, Romania
*email: cfarcau@itim-cj.ro
ORCID: https://orcid.org/0000-0001-8602-0771

**Abstract:** Colloidal plasmonic-photonic crystals represent a class of hybrid materials composed of a dielectric colloidal spheres photonic lattice and a metal plasmonic film. In this work, the optical properties of a linear array colloidal plasmonic-photonic crystal consisting of silver films deposited over linear arrays of polystyrene microspheres are analysed in detail. Experimental and simulated optical transmittance and reflectance spectra both with unpolarized and polarized light are used to investigate the optical response of the linear plasmonic-photonic crystal. Among the various photonic/plasmonic modes observed, the existence of both propagative plasmonic-photonic hybrid mode and localized surface plasmon mode can be mentioned. The spectral tunability of these structures is highlighted by studying the dependence of the optical response on geometrical parameters such as sphere diameter and grating period. Finally, the linear plasmonic-photonic crystal exhibits a polarization-selective surface-enhanced Raman scattering effect, making them of interest for both fundamental studies and development of applications based on surface-enhanced Raman spectroscopy or surface-enhanced fluorescence.

## Statements and Declarations

Funding: Partial financial support was received from the Romanian Ministry of Education and Research, CNCS-UEFISCDI, project number PN-III-P4-ID-PCE-2020-1607, within PNCDI III.

Competing Interests: The authors have no competing interests to declare that are relevant to the content of this article.

## 1. Introduction

Hybrid plasmonic-photonic crystals are periodically structured metal-dielectric structures which offer attractive ways of manipulating optical waves at the micro and nano scale[1–3]. They generally consist of a dielectric photonic crystal superposed (or underposed) to a metallic nanoparticle or thin film which can support localized or propagating surface plasmon modes. The close vicinity of the dielectric and metallic grating implies a strong interaction between the modes characteristic to each component, and this leads to a distinct coupled, or hybrid regime. As an example, an interesting case results when a surface plasmon resonance matches the wavelength of the photonic bandgap provided by a photonic crystal, leading to the excitation of a particular resonance known as Tamm plasmon mode[4]. Colloidal plasmonic-photonic crystals (CPPC) are a class of such hybrid materials in which the dielectric photonic lattice is made of colloidal crystals, i.e. periodic 2D or 3D arrangements of dielectric, usually polystyrene or $SiO_2$, colloidal spheres. These can be assembled directly over metallic films or can be coated by a metal film which covers the top colloidal layer, being thus periodically corrugated[3,5]. The involved fabrication procedures are parallel, faster, and simpler compared to classical lithographic methods such as electron-beam or focused ion-beam lithography. The colloidal crystals can be obtained from aqueous colloidal suspensions by different self-assembly strategies, while metal films can be deposited by thermal or e-beam evaporation or magnetron sputtering. Besides the relative ease of fabrication, another useful feature of CPPC is the possibility of engineering the optical response by the colloidal sphere diameter and composition, and the thickness and composition of the metal film.

Among the CPPC, metal-coated colloidal crystal monolayers are ones of the most widespread, their optical properties being now pretty well understood. They have been attractive plasmonic-photonic materials because they offered an easy to prepare platform for studying fundamental plasmonic aspects, but also due to their high application potential. Several groups have highlighted the exciting optical properties possessed by this type of structures, such as, an enhanced light transmission behaviour[6–10] similar with the enhanced or extraordinary optical transmission (EOT)[11]. Other authors demonstrated how strong coupling between the Bragg-plasmon mode supported by the CPPC and molecular excitons of J-aggregates covering the hybrid structure can be manipulated[12]. Noble metal-coated colloidal monolayers have proved their utility as platforms for enhanced optical spectroscopy applications such as Surface-Enhanced Raman Spectroscopy (SERS) [13–15] and Surface-Enhanced Fluorescence[16–18]. One of the drawbacks in the fabrication process is the difficulty of obtaining large areas of defect free monolayers of colloids. A typical result of a self-assembly process is a hexagonal close packed lattice presenting a series of defects like vacancies, cracks and domains with different crystal orientations. There are few examples in the literature reporting methods to fabricate large areas of hexagonal close packed colloidal monolayers with few defects. As example, by adjusting parameters of spin-coating mediated by surfactants large surfaces could be covered [19]. Assembly on a water-air interface has also evolved, in a recent report acoustic waves being used to improve colloidal crystal quality [20]. One approach to better control crystal orientation is by using patterned substrates, which can guide the colloidal self-assembly process. As an example, Banik and Mukherjee showed that linear colloidal arrays can be obtained by spin-coating on a grating-like substrate, by adjusting colloid concentration and rotation speed[21].We have previously reported on a method to obtain linear colloidal arrays by convective self-assembly (CSA) on DVD templates[22], and used these to fabricate an original class of hybrid metal-dielectric periodic structure consisting in linear array metal-coated microspheres (LA-MCM). The structure we proposed has a well-controlled crystal orientation and preliminary optical investigations promise potential applications in the field of plasmonics.

In this work, the optical properties of the linear colloidal plasmonic-photonic crystal consisting of silver films deposited over linear arrays of polystyrene microspheres are analysed in detail. Both experiments and simulations are employed, aiming at disentangling the various optical/plasmonic modes involved, and their dependence on geometrical parameters. Optical transmission and reflectance measurements both with unpolarized and polarized light were used to investigate the plasmonic response of the LA-MCM. Finite-Difference Time-Domain simulations are employed in order to identify and understand the nature of the different plasmon modes excited in this hybrid metal-dielectric structure, and highlight their dependence on microsphere diameter and grating period. It is also demonstrated that the linear plasmonic-photonic crystal exhibits a polarization-selective surface-enhanced Raman scattering effect, making them of interest for both fundamental plasmonics studies and development of applications based on surface-enhanced Raman spectroscopy or surface-enhanced fluorescence.

## 2. Experimental details

*2.1. Fabrication.* Linear Array Colloidal Crystals (LA-CC) were prepared by convective self-assembly (CSA) on the patterned inner surface of a Digital Versatile Disc (DVD). The pre-treatment of the DVD surface was described in detail previously [22]. Briefly, it involves mechanical separation of the two sandwiched polycarbonate discs, their cleaning, rinsing, drying followed by UV-ozone treatment to render the surface hydrophilic. Aqueous suspensions (2% W/v) of plain polystyrene spheres (Microparticles GmbH) with 497 nm nominal diameter were used. These were adjusted to about 0.33% W/v by dilution with distilled water. For CSA, a home-made setup, as schematized in Figure 1a was used.

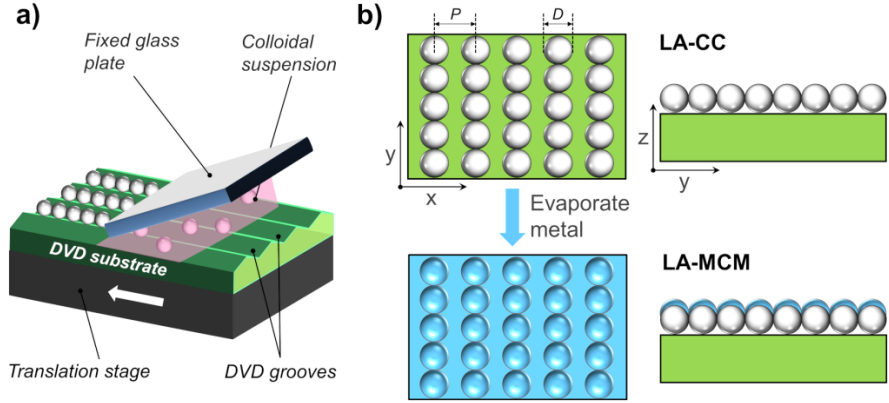

**Fig. 1**(a) Scheme of the CSA process on DVD substrate; b) Schemes of the Linear Array Colloidal Crystals (LA-CC) and Linear Array Metal-Coated Microspheres (LA-MCM)

The DVD substrate was fixed on a motorized linear translation stage, controlled through computer software. A rectangular glass plate was attached to a pole at an angle of about 25º, such that one of its edges faces the substrate parallel to it and perpendicular to the DVD grooves. A drop of colloid (5 - 10 µl) was inserted into the wedge formed by the glass plate and the DVD substrate. The stage was then translated relative to the fixed glass plate, at speeds ranging between 4-7µm/s, along a direction parallel to the substrate channels (see Figure 1a). The process took place in ambient conditions (30-32$^0$C and 28-34%RH). A 50 nm thick silver (Ag) film was then deposited on the prepared LA-CC by means of vacuum thermal evaporation at normal incidence. Film thickness was monitored by a quartz crystal microbalance.

*2.2. Morphological and optical characterisation.* To characterize the samples, optical microscopy, Scanning Electron Microscopy (SEM) on a Quanta 3D FEG - FEI and Atomic Force Microscopy (AFM) on a WITec system were employed. Optical measurements were made with an Ocean Optics USB4000 UV-VIS spectrometer. For transmission measurements the spectrometer was connected to an inverted Olympus optical microscope, and for reflection measurements an upright Zeiss microscope was used. In both cases a 10X objective was used to illuminate or collect light, which assures a negligible amount of scattered/diffracted light being collected. A 0.6 mm core diameter optical fibre was used to transmit the collected light to the spectrometer, which reduces the area of the sample being analysed to a disk of about 60 µm in diameter. For measurements in polarized light a polarizer was added in the setup, such that the light reaching the sample had the desired polarization.

*2.3. FDTD simulations.* The interaction between electromagnetic radiation and the LA-CC or LA-MCM was simulated using the Ansys Lumerical FDTD software. A monolayer of dielectric spheres are arranged in a linear array structure, on a dielectric substrate. A finite array composed of 90 dielectric spheres was simulated. The sphere array was covered with a silver film. Silver particles formed on the substrate (as in real experiments) are also considered. Material properties used are those from Lumerical's database, based on Palik Handbook of Optical Constants of

Solids for SiO$_2$ (glass substrate) and CRC Handbook of Chemistry & Physics for Ag. Refractive indexes of 1.59 and 1 were used for polystyrene and air, respectively. The Ag film deposited on top of the polystyrene spheres was represented as ellipsoidal caps. The detailed definition of these ellipsoidal caps was presented elsewhere [9]. In simulations presented here we used dtF=40 and dzF=50. Perfectly matched layer (PML) boundary conditions were used for all boundaries. A mesh accuracy factor of 6 was used and a conformal variant 1 mesh refinement. Throughout the metal structure and the surrounding area (50 nm above and below), a finer mesh was defined, with a spatial resolution of 5 nm, a factor 100 smaller than the diameter of the spheres. A plane wave source, linearly polarized along the X or Y direction was used, spanning the spectral range 450-950 nm.

## 3. Results and discussion

Figure 2a presents a typical SEM image of prepared LA-CC. The alignment of the colloids during the convective self-assembly process is a result of confinement between the water meniscus and the DVD grating surface. As the meniscus positioned perpendicular to the grating is translated along the grating lines, the spheres are trapped and deposit inside the DVD grooves. Some defects are also found such as interrupted chains, or larger spheres perturbing the array. The 3D rendering of the AFM image in Figure 2b better highlights the topography of the obtained LA-MCM.

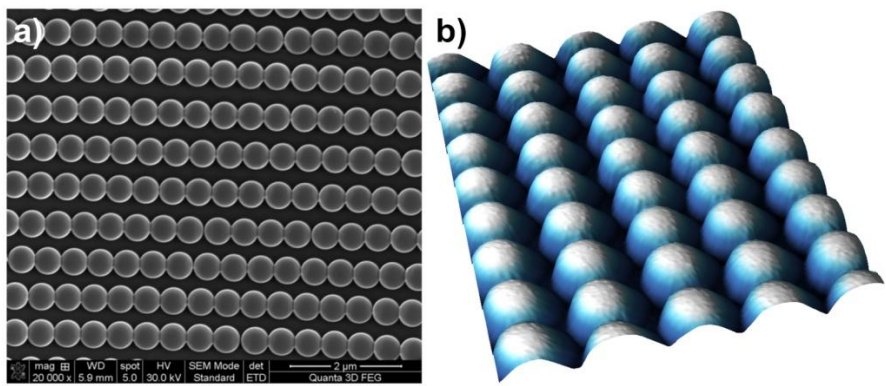

**Fig. 2**(a) SEM image of linear array colloidal crystal;(b) AFM 3D rendering of linear array metal-coated microspheres

The metal film on top of the dielectric spheres forms a network of interconnected metal caps or half-shells. Long, continuous metal strips are formed on the substrate, due to metal deposited in the empty space between the chains of spheres.

## 3.1. Overall optical response of LA-MCM

Next, we analyse the optical response of both LA-CC and LA-MCM in comparison with the better-understood optical response of standard, hexagonal lattice colloidal crystals (hex-CC), and metal-coated microsphere arrays based on those (hex-MCM). The unpolarized transmittance and reflectance spectra of LA-CC are presented in Figure 3(a).

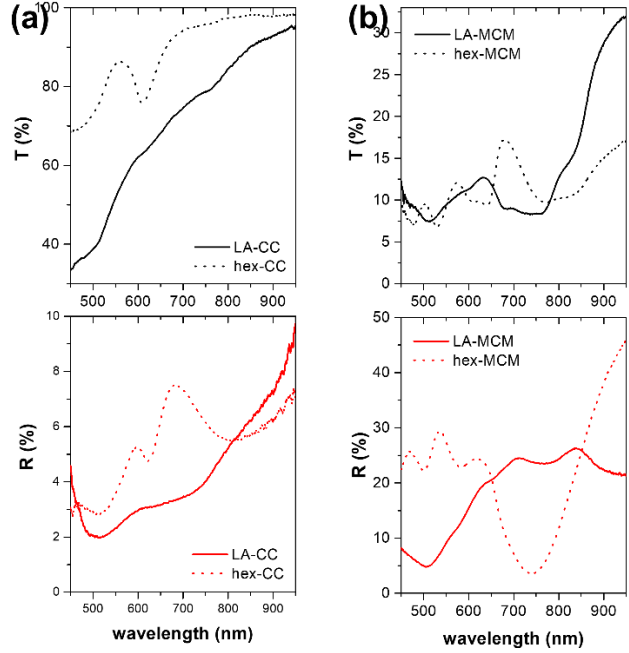

**Fig. 3** (a) *T* and *R* spectra of linear array colloidal crystals (LA-CC) compared to standard hexagonal array colloidal crystals (hex-CC); (b) *T* and *R* spectra of linear array metal-coated microspheres (LA-MCM) compared to standard hexagonal array metal-coated microspheres (hex-MCM)

First let us note that the monolayer hexagonal CC (hex-CC) is characterized by a dip in its transmittance spectrum (*T*), observed at 610 nm in Figure 3a. Since bulk polystyrene is transparent at these wavelengths, this dip in transmission represents a specific optical response of ordered 2D arrays of dielectric microspheres. At this wavelength, photons are coupled into eigenmodes of the spheres lattice, which are responsible for their guided propagation in the plane of the dielectric spheres, along chains of spheres [23]. As such, we can refer to this channel of energy dissipation responsible for the minima in the measured transmittance spectra as the first-order photonic guided mode or resonance. As a first approximation, the wavelength of this mode can be estimated by $\lambda_{res} = 2\pi n_{ef}/G_{ij}$, where $n_{ef}$ is an effective refractive index of the sphere array (containing also air) and $G_{ij} = (4\pi/\sqrt{3}D)\sqrt{i^2 + j^2 + ij}$ is a reciprocal vector of the array. At this spectral position the grating becomes a resonant structure and the diffracted orders become evanescent, to some extent similar to Wood's anomaly [24]. The calculation yields a value of 590 nm for $\lambda_{res}$ value, which is blue-shifted with respect to the position of the experimental transmittance dip, probably due to the presence of the substrate, and other geometrical factors impacting the precise estimation of $n_{ef}$. The wavelength of the guided resonance transmittance minima thus scale with the diameter of the dielectric spheres, while their depth and width are a good indicator of the degree of order of the CC. The decreasing trend in the spectrum, as approaching the UV region, is due to the increasing absorption of light by polystyrene and presence of several additional higher order modes, which are close to each other and overlap. The reflectance (*R*) spectrum of the hex-CC is characterized again by a dip attributed to the same mode as described above. This mode thus manifests itself both in *T* and *R* spectra, representing an amount of the incident photon flux that is not transmitted, is not reflected, but is coupled into eigenmodes of the photonic lattice, and propagates in the plane of the lattice. This mode is barely visible in the *T* and *R* spectra of the LA-CC, but its origin is the one discussed above. Because there is only one axis along which photons can couple, compared to three axes for the hex-CC, this less efficient coupling can be understood.

The *T* spectrum of the hex-MCM (dotted lines in Fig. 3b) is dominated by a transmission band peaking at 680 nm, which has been analyzed before[7], and attributed to a phenomenon similar to the Enhanced Optical Transmission (EOT) reported for nanohole arrays in metallic films[11]. This transmission band, in a spectral region in which a flat film of same thickness is opaque, is a result of excitation of propagative surface plasmon modes in the metal film mediated by the photonic modes in the underlying dielectric sphere lattice. In fact, the position of the guided-mode resonance in the hex-CC always coincides to the local minimum at the blue side of the EOT-like maximum. In plasmonic crystals, this minimum is associated to light diffracted parallel to the lattice surface, and is known as Wood anomaly or Rayleigh–Wood anomaly[24,25]. It indicates the wavelength above which free space light diffraction is forbidden for the given order. The *R* spectrum of the hex-MCM is dominated by a broad dip with a minimum located at 740 nm. This coincides with the maximum absorbance in the metal film, attributed to excitation of localized surface plasmons at the junctions between neighbouring metal half-shells.

The optical response of the LA-MCM apparently deviates rather strongly from the response of the hex-MCM. Only a very low-intensity T band can be observed for the LA-MCM at 696 nm, near the wavelength of hex-MCM EOT band, while a stronger transmission is observed at shorter wavelength (633 nm). The R spectrum of LA-MCM is also quite different from the hex-MCM, with no well-defined minimum around the 740 nm. It is therefore necessary to investigate the optical T and R behaviour in polarized light, in order to understand their plasmonic/photonic behaviour.

## 3.2. Polarization-sensitive optical response

This section starts with the analysis of the polarized optical response of LA-CC. The incident light was polarized either parallel (00°, $Y$ direction in Figure 1b) or perpendicular (90°, $X$ direction in Figure 1b) to the linear chains of metal-coated spheres. Both experimental and simulated T, R, and inferred A spectra are presented in Figure 4 for the two polarizations (00° - parallel, 90° - perpendicular).

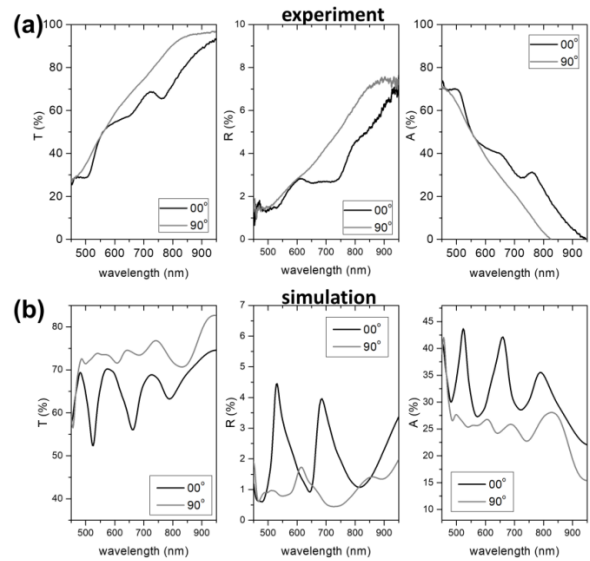

**Fig. 4** (a) Experimental polarized $T$, $R$, and $A$ spectra of LA-CC; (b) Simulated polarized $T$, $R$, and $A$ spectra of LA-CC

The stronger optical interactions are observed for the parallel polarization. T spectra exhibit three minima in the analyzed range, both in experiments (500 nm, 650 nm, and 760 nm) and simulations (524 nm, 662 nm, and 787 nm), for light polarized along the sphere chains. The R spectra also present some minima, although of much smaller amplitude. The inferred absorbance ($A$) in turn show bands peaking around 500 nm, 650 nm, and 761 nm in experimental spectra, respectively at 522 nm, 657 nm, and 790 nm in simulations. Although intensities and band widths obviously differ between experiment and simulation, a good overall agreement between the two can be observed for spectral shape and band positions. Note also that the mentioned $A$ bands are not due to actual absorbance, but due to photonic effects of coupling to the guided modes along chains of spheres.

The polarized optical response of LA-MCM is next presented in Figure 5. For light polarized parallel to the structure's main axis (00°, $Y$ direction), a $T$ band can be observed at about 705 nm. Its position to the long-wavelength side of the dip at 650 nm from the $T$ spectrum of the LA-CC. This correspondence suggests the same situation as for the EOT-like $T$ band in hex-MCM. By analyzing the electric field components at 664 nm, at the simulated $T$ band maximum (Figure 5c, and symbol ♦ in Figure 5b), one can notice the same field distribution as previously observed for hex-MCM [8]. An important $E_Z$ field component can be observed, with the lobes extending over the metal coating and the dielectric spheres, this Z-field indicating light propagating along the Y direction. As such, this is a *T* band involving excitation of propagative surface plasmons coupled to the photonic guided resonance of the dielectric sphere lattice.

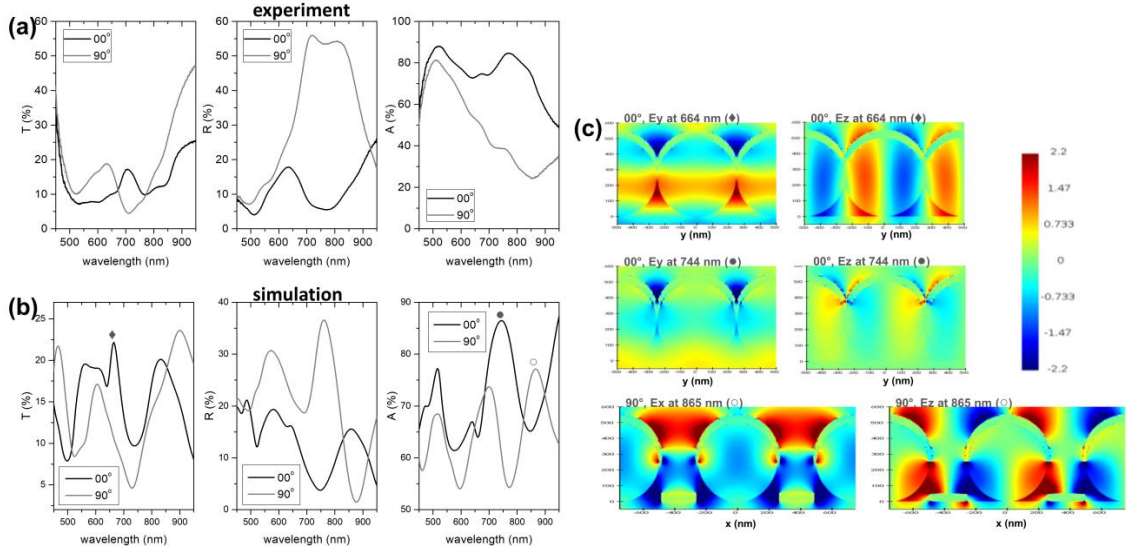

**Fig. 5** (a) Experimental polarized *T*, *R*, and *A* spectra of LA-MCM; (b) Simulated polarized *T*, *R*, and *A* spectra of LA-MCM; (c) Electric field distribution at selected wavelengths, indicated by symbols on panel (b)

A thing to notice concerning reflectance spectra is that the LA-MCM exhibits a strong contrast between the two polarizations in the range 650 nm - 850 nm. By next analyzing also the inferred *A* spectra, it can be observed that these also present a rather strong contrast between the two polarizations in this range. In fact, for the parallel polarization (00°) the maximum absorbance coincides with the minimum reflectance. The electric field components at 744 nm, at the simulated *A* band maximum (Figure 5c, and symbol ● in Figure 5b), indicate a strong localization of the $E_Y$ component at the junction between metal half-shells. Since the $E_Z$ component is also localized at the same location, and less intense, it indicates that in this case, a localized plasmon resonance is excited. For the perpendicular polarization (90°, *X* direction) some absorbance bands can also be observed, less intense than for the parallel polarization. The *A* band at 865 nm in the simulated spectrum (Figure 5c, and symbol ○ in Figure 5b) also coincides with a *R* minimum, and presents an interesting electric field distribution. Both the $E_X$ and $E_Z$ electric fields components have considerable magnitudes not only on the metallic half-shells but also on the metal strips on the substrate. Further analyses, beyond the scope of this work, are needed to fully understand the role of these long metal strips formed on the substrate.

## 3.3. Spectral tunability by sphere size and grating period

Next, the influence of sphere size ($D$) and period of the grating ($P$) on the optical response of LA-MCM was explored, and results are displayed on Figure 6. A first observation is that the dependence on sphere size is visible in the $T$, $R$, and $A$ spectra for both polarizations, sphere size having an impact on most of the spectral features. The general trend is that increasing the sphere size induces redshifts on band maxima and dip minima spectral positions.

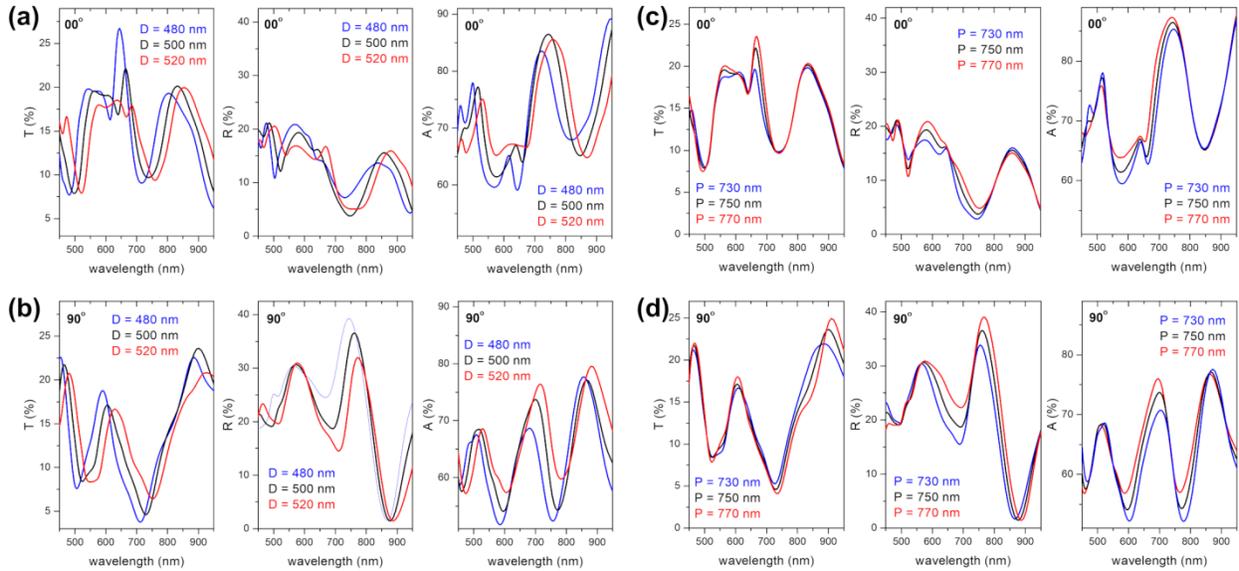

**Fig. 6** Sphere size ($D$) dependence of $T$, $R$, and $A$ spectra for LA-MCM polarized parallel (a) and perpendicular (b) to the LA-MCM axis. Grating period ($P$) dependence of $T$, $R$, and $A$ spectra for LA-MCM polarized parallel (c) and perpendicular (d) to the LA-MCM axis.

The effect of grating period is overall less pronounced, with stronger modifications being observed for the perpendicular polarization. These results indicate that by tuning the sphere size and grating period the optical response of the LA-MCM can be tuned, with a practical impact as it allows to tune also the polarization contrast of this anisotropic plasmonic-photonic nanostructure.

## 3.4. Polarization sensitivity in SERS

Finally, some SERS results are presented, as an example of application exploiting the polarization-sensitive optical/plasmonic interactions in LA-MCM. As such the SERS spectra of a layer of cresyl violet molecules deposited on top of the silver film were recorded with the excitation laser polarization being parallel or perpendicular to the LA-MCM axis. 785 nm was selected as excitation wavelength, as it falls in the spectral region where the structure shows a strong polarization contrast for the $A$ spectra (Figure 5a). In order to avoid any errors due to equipment components polarization sensitivity, the polarized excitation was kept fixed, and the sample was rotated by 90°. Figure 7 presents the SERS spectra corresponding to the two excitation configurations.

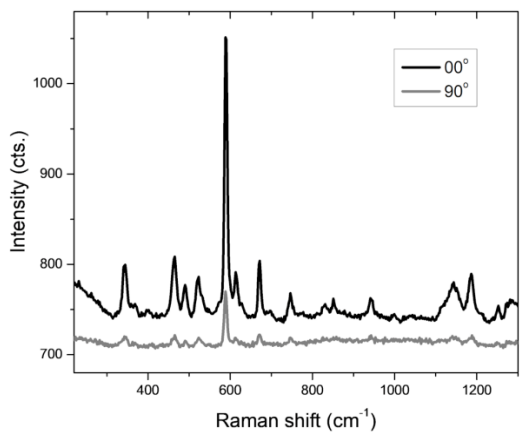

**Fig. 7** SERS spectra of cresyl violet adsorbed on LA-MCM under polarized excitation at 785 nm

The difference in intensity between the two SERS spectra is obvious, with the parallel polarization yielding 5-6 times more intense SERS bands. For the parallel polarization, the electric fields are strongly enhanced at the junctions between aligned metal half-shells, leading to higher SERS enhancements. Such polarization-selective SERS substrates could prove useful for developing SERS applications as they would allow a good control over the precise location of the highest SERS-enhancing surface locations. Moreover, by fabricating such linear array plasmonic crystals, mechanically adjustable plasmon resonances could be achieved, which in turn can serve to adjust in real-time the electromagnetic enhancements in surface-enhanced Raman scattering or surface-enhanced fluorescence.

## 4. Conclusions

Linear array metal-coated microspheres (LA-MCM) were fabricated by depositing silver films over linear array colloidal crystals fabricated by convective self-assembly on a DVD grating substrate. The polarized optical response of the LA-MCM was analyzed both experimentally and by FDTD simulations in the range 450 nm - 950 nm. For light polarized parallel to the structure's main axis, an EOT-like transmittance band was observed, involving excitation of propagative surface plasmons coupled to the photonic guided resonance of the dielectric sphere lattice. Reflectance spectra of LA-MCM exhibit a strong contrast between the two polarizations, which can be observed also for absorbance spectra. For the parallel polarization, the maximum absorbance coincides with the minimum reflectance, electric fields of this mode being mainly polarized along the LA-MCM axis. For the perpendicular polarization some less intense absorbance bands can also be observed, presenting an interesting electric field distribution, with $E_X$ and $E_Z$ electric fields components of considerable magnitudes both on the metallic half-shells and on the metal strips on the substrate. The sphere size has an impact on most of the spectral features in the spectra for both polarizations, the general trend being that increasing the sphere size induces redshifts of band maxima and dip minima. The effect of grating period is less pronounced, with stronger modifications being observed for the perpendicular polarization. SERS measurements indicate 5-6 times more intense SERS spectra recorded with the excitation polarized parallel to the LA-MCM axis. The results can prove useful for the development of plasmonic components such as polarizers or for developing niche applications based on surface-enhanced Raman scattering or surface-enhanced fluorescence.

**Acknowledgement.** CF acknowledges support by a grant of the Romanian Ministry of Education and Research, CNCS-UEFISCDI, project number PN-III-P4-ID-PCE-2020-1607, within PNCDI III. V. Saracut is acknowledged for contributing to the fabrication of the colloidal uniaxial arrays. A. Vulpoi is acknowledged for support on SEM imaging.

**Data Availability Statement**. The datasets generated during and/or analysed during the current study are available from the corresponding author on reasonable request.

**Graphical Abstract.**

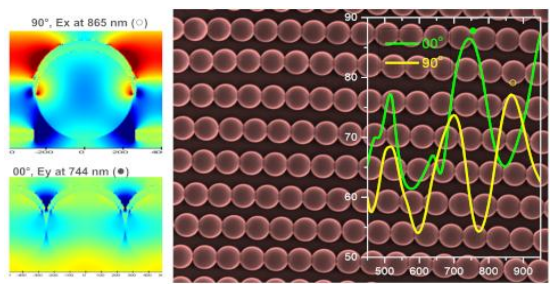